\PassOptionsToPackage{bookmarks=false}{hyperref}
\documentclass[sigconf]{acmart}

\usepackage{multirow}
\usepackage{amsmath}
\usepackage{listings}
\usepackage{xcolor}

\definecolor{dkgreen}{rgb}{0,0.6,0}
\definecolor{gray}{rgb}{0.5,0.5,0.5}
\definecolor{mauve}{rgb}{0.58,0,0.82}

\AtBeginDocument{%
  \providecommand\BibTeX{{%
    \normalfont B\kern-0.5em{\scshape i\kern-0.25em b}\kern-0.8em\TeX}}}

\copyrightyear{2020}
\acmYear{2020}
\setcopyright{acmcopyright}\acmConference[ASE '20]{35th IEEE/ACM International Conference on Automated Software Engineering}{September 21--25, 2020}{Virtual Event, Australia}
\acmBooktitle{35th IEEE/ACM International Conference on Automated Software Engineering (ASE '20), September 21--25, 2020, Virtual Event, Australia}
\acmPrice{15.00}
\acmDOI{10.1145/3324884.3416591}
\acmISBN{978-1-4503-6768-4/20/09}

\begin{document}

\title{Multi-task Learning based Pre-trained Language Model for Code Completion}


\author{Fang Liu}
\affiliation{%
  \institution{Key Lab of High Confidence Software \\ Technology, MoE (Peking University)\\ Beijing, China}
}
\email{liufang816@pku.edu.cn}

\author{Ge Li}
\authornote{Corresponding authors.}
\affiliation{%
  \institution{Key Lab of High Confidence Software \\ Technology, MoE (Peking University)\\ Beijing, China}
}
\email{lige@pku.edu.cn}

\author{Yunfei Zhao}
\affiliation{%
  \institution{Key Lab of High Confidence Software \\ Technology, MoE (Peking University)\\ Beijing, China}
}
\email{zhaoyunfei@pku.edu.cn}

\author{Zhi Jin}
\authornotemark[1]
\affiliation{%
  \institution{Key Lab of High Confidence Software \\ Technology, MoE (Peking University)\\ Beijing, China}
}
\email{zhijin@pku.edu.cn}


\renewcommand{\shortauthors}{Liu et al.}

\begin{abstract}
Code completion is one of the most useful features in the Integrated Development Environments (IDEs), which can accelerate software development by suggesting the next probable token based on the contextual code in real-time. Recent studies have shown that statistical language modeling techniques can improve the performance of code completion tools through learning from large-scale software repositories. However, these models suffer from two major drawbacks: a) Existing research uses static embeddings, which map a word to the same vector regardless of its context. The differences in the meaning of a token in varying contexts are lost when each token is associated with a single representation; b) Existing language model based code completion models perform poor on completing identifiers, and the type information of the identifiers is ignored in most of these models. To address these challenges, in this paper, we develop a multi-task learning based pre-trained language model for code understanding and code generation with a Transformer-based neural architecture. We pre-train it with hybrid objective functions that incorporate both code understanding and code generation tasks. Then we fine-tune the pre-trained model on code completion. During the completion, our model does not directly predict the next token. Instead, we adopt multi-task learning to predict the token and its type jointly and utilize the predicted type to assist the token prediction. Experiments results on two real-world datasets demonstrate the effectiveness of our model when compared with state-of-the-art methods.
\end{abstract}



\maketitle

\section{Introduction}
As the complexity and scale of the software development continue to grow, large corpora of open source software projects present an opportunity for modeling source code on machine learning \cite{allamanis2018survey}. Most of these approaches are based on the observation of source code's naturalness \cite{hindle2012naturalness}, that is, source code is written by humans and for humans to read, it displays some of the statistical properties as natural language. Thus, statistical language models have been used for source code modeling \cite{hindle2012naturalness,tu2014localness,raychev2014code}, benefiting many software engineering tasks, including code summarization \cite{hu2018deep,wan2018improving}, code clone detection \cite{wei2017supervised,zhang2019novel} program repair \cite{gupta2017deepfix,vasic2019neural}, especially, in code completion \cite{hindle2012naturalness,tu2014localness,hellendoorn2017deep,Li2018Code}.

Code completion is an essential feature of Integrated Development Environments (IDEs). It speeds up the process of software development by suggesting the next probable token based on existing code. In recent years, as the success of deep learning, Recurrent Neural Network (RNN)-based language models have been applied to source code modeling \cite{bhoopchand2016learning,Li2018Code}. In these models, a piece of source code is represented as a source code token sequence or an Abstract Syntactic Tree (AST) node sequence. Given a partial code sequence, the model computes the probability of the next token or AST node and recommends the one with the highest probability. Furthermore, these language models can also learn useful word embeddings, which can be used for other downstream tasks in the same way as word2vec-style embeddings \cite{mikolov2013distributed}. However, source code has some special properties, which have not been exploited in existing statistical language models. We discuss two critical issues in detail below.

\lstdefinestyle{interfaces}{
  float=tp,
  floatplacement=tbp,
  abovecaptionskip=0pt
}

\begin{lstlisting}[style=interfaces,caption={A Java method example.}]
public long getMaximumTime(IoEventType type) {
    if (!timerManager.containsKey(type))
        throw new IllegalArgumentException("Please add this event first.");
    return timerManager.get(type).getMaximum();
}
\end{lstlisting}\label{java_m}

The contextual information is not well considered in the existing code completion models. Writing clean and readable code that conforms to the specification has been paid more attention in software development, which helps the developers reuse and maintain the code. When programming, developers tend to use meaningful and conventional identifier names and natural language documentation \cite{Martin2009}. As a result, information contained in the source code can be exploited by machine learning algorithms. Most of these models are based on learned representations called embeddings, which transform words into a continuous vector space \cite{mikolov2013distributed}. However, existing research \cite{bhoopchand2016learning,Li2018Code,karampatsis2020big} uses static embeddings, which map a word to the same vector regardless of its context. For example, in Java method overloading, the same function name can have different meanings based on the number and type of the parameters. However, the static embedding will map it to the same vector. The differences in the meaning of a token in varying contexts are lost when each token is associated with a single representation. The surrounding tokens of the program entities usually contain certain information that reflects the roles of the entities. For instance, for a method name, the surrounding tokens might include the variables/fields/methods that are used/accessed/invoked to implement the method. Taking the Java method in Code \ref{java_m} as an example, the function name \textit{getMaximumTime} can be inferred from the variables’ names and method calls in the body, e.g., \textit{getMaximum}, \textit{timerManager}. These tokens provide information about possible values the function could take, and so should affect its representation. 

\begin{figure}[t] 
\setlength{\abovecaptionskip}{0cm}
\centering\includegraphics[width=8.5cm]{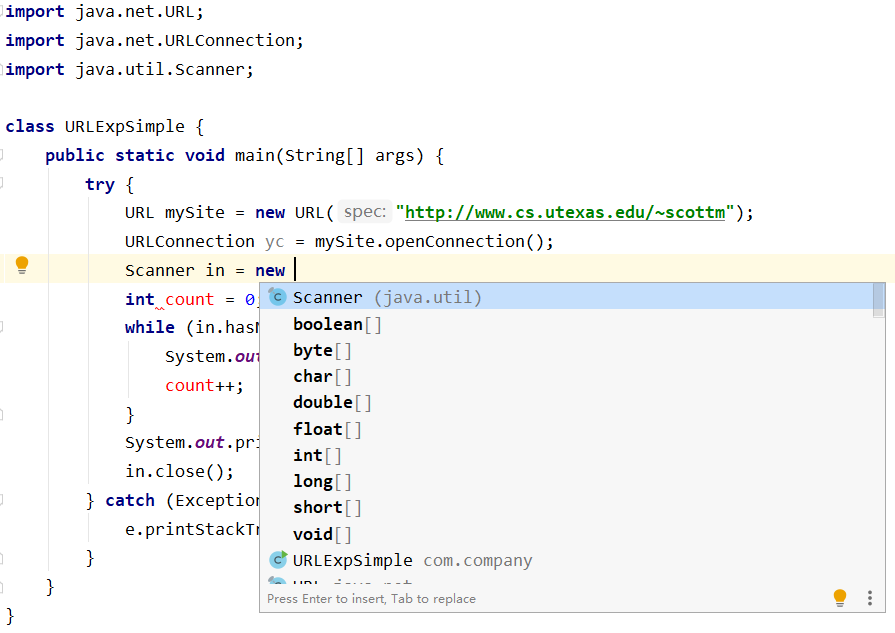} \caption{Java IDE completion example.}
\label{Fig:java_ide}
\vspace{-0.3cm}
\end{figure}

Identifier completions are challenging, and existing statistical Language Model (LM) based code completion models perform poorly on completing identifiers. These approaches consider every token in the source code file as targets for completion. More than two-thirds of the completions do not refer to identifiers. Instead, the majority concern punctuation-like tokens (e.g., operators, braces), which are much easier to complete than identifiers, but these completions are not that beneficial to developers \cite{karampatsis2020big}. Besides, the type information of the identifiers is ignored in most of the models. Modern IDEs for most languages heavily rely on types to make helpful suggestions for completing partial code. For example, when accessing the field of an object in a Java IDE, code completion suggests suitable field names based on the object’s type \cite{malik2019nl2type}. Taking the code completion example of a Java IDE ( ItelliJ IDEA) in Figure \ref{Fig:java_ide} as an example, the IDE suggests ``Scanner'' as the next token based on its type (i.e., java.util), and not just predict the frequent token in the corpus. For those dynamic languages, such as Python and JavaScript, IDEs often fail to make accurate suggestions because the types of code elements are unknown, which further demonstrates the importance of the type information. However, most of the existing LM-based source code modeling techniques and code completion studies do not take the type information into consideration.

In response to the observations and concerns raised above, we have developed a Code Understanding and Generation pre-trained Language Model (CugLM) for source code modeling. Recent work on pre-trained language models has found that the contextual embeddings produced by these models can lead to better performance for many natural language processing (NLP) tasks \cite{Devlin2019Bert,Peters2018Deep,radford2018improving,Howard2018Universal}. In these models, the representation for each word is learned using the language models, where the vector of the word is computed based on the context it is used. Thus, the vector of the same word under different contexts can be different.  In particular, BERT \cite{Devlin2019Bert} proposes a bidirectional Transformer Encoder with two new pre-training objective: ``masked language model'' and ``next sentence prediction'', where ``masked language model'' randomly masks some of the tokens from the input, and the objective is to predict the masked word based only on its context, and ``next sentence prediction'' predicts whether two sentences follow each other in a natural discourse. By using these two objectives, BERT can produce powerful bidirectional contextual representations and advances the state-of-the-art for many NLP tasks. Inspired by the success of pre-trained language models in NLP, we propose a multi-task learning based pre-trained language model to produce general and contextual representations for programs that can broadly support code understanding and generation tasks, and then apply it to code completion. During the pre-training period, we adopt the multi-task learning framework to learn the following three training objectives jointly: 

\textbf{1) Masked bidirectional Language Modeling}: The identifiers are more informative for understanding the program and correctly suggesting the identifiers is challenging in existing code completion research \cite{karampatsis2020big}. Thus, producing contextual and general representations for tokens, especially for identifiers, would be helpful for source code modeling and code completion. For these reasons, we mask the identifiers from the programs, and the objective is to predict the masked tokens based on their bidirectional context.

\textbf{2) Next Code segment Predicting}: We argue that understanding relationships between code segments can help in source code modeling. In order to achieve this, we pre-train a binarized next code segment prediction task, that is, predicting whether two segments of code tokens follow each other in a piece of code snippet.

\textbf{3) Unidirectional Language Modeling}: a left-to-right language modeling task, where the representation of each token encodes only the leftward context tokens and itself. This training objective is added because for the generation tasks (e.g., code completion), only leftward contextual tokens are allowed. 

After the model has been pre-trained, we fine-tune it (directly apply the pre-trained model and adapt the model on downstream tasks by fine-tuning the pre-trained parameters) on the code completion task. During the code completion, our model does not directly predict next token, instead, we adopt a multi-task learning framework to predict the token and its type. We first predict the type of the token, and then use predicted type to assist the token prediction. 

We create two massive corpora of Java and TypeScript programs collected from GitHub to pre-train and fine-tune the model. We compare our model with two state-of-the-art code completion approaches: Byte Pair Encoding based Neural Language Model (BPE NLM) \cite{karampatsis2020big} and Pointer Mixture Network \cite{Li2018Code}. For completing all types of tokens, our model achieves the accuracy of 80\% and 81\% on Java and TypeScript datasets, respectively, which improves Pointer Mixture Network by 17\% and 24\%, and improves BPE NLM by 19\% and 24\%, in terms of relative improvements. For identifier completion, our model achieves the accuracy of 48\% and 39\%, respectively, which improves Pointer Mixture Network by 29\% and 34\%, and improves BPE NLM by 11\% and 9\%, in terms of relative improvements. 

The main contributions of this paper are summarized as follows:
\begin{itemize}
    \item We present the first attempt at pre-training a language model with a transformer-based architecture for code completion.
    \item We take advantage of the type information to help our model make better suggestions on identifiers.
    \item We compare our model with state-of-the-art code completion models and evaluate the performance of these models on two real-world datasets. Experimental results demonstrate that our model achieves the best performance compared with the baseline models. 
\end{itemize}

\section{Background}\label{background}
\subsection{Statistical Language Model}
Statistical language models capture the statistical patterns in languages by assigning occurrence probabilities to a sequence of words in a particular sequence, which will score an utterance high, if it sounds ``natural'' to a native speaker, and score low the unnatural (or wrong) sentences. Programming languages are kind of languages that contain predictable statistical properties \cite{hindle2012naturalness}, which can be modeled by statistical language models. Given a token sequence S = $s_{1},s_{2},...,s_{t}$, the probability of the sequence is computed as: 
\begin{equation}
p(S)=p(s_1)p(s_{2}|s_{1})p(s_{3}|s_{1}s_{2}),...,p(s_{t}|s_{1}s_{2},...,s_{t-1})
\end{equation}

The probabilities are hard to estimate when the number of the context tokens $s_{1},s_{2},...,s_{t-1}$ is tremendous. The N-gram model based on the Markov assumption is proposed to address this challenge, where the probability of a token is dependent only on the $n-1$ most recent tokens. N-gram based models have been generally applied to code completion \cite{hindle2012naturalness,tu2014localness,hellendoorn2017deep}. These models have been proved to capture the repetitive regularities in the source code effectively. In recent years, deep recurrent neural networks, including Long Short-Term Memory (LSTM) \cite{hochreiter1997long} and Gate Recurrent Unit (GRU) \cite{cho2014properties}, have shown great performance on modeling programming languages \cite{liu2016neural,bhoopchand2016learning,Li2018Code}. By using recurrent connections and gate mechanisms, information can cycle inside these networks for a long time, which loosens the fixed context size and can capture longer dependencies than the N-gram model. 

However, the introduction of the gating mechanism in LSTMs and GRUs might not be sufficient to address the gradient vanishing and explosion issue fully. To ease this issue, attention mechanisms \cite{bahdanau2014neural,vaswani2017attention}, which add direct connections between long-distance word pairs, are proposed. For example, the Transformer \cite{vaswani2017attention} is an architecture based solely on attention mechanism. It uses a multi-headed self-attention mechanism to replace the recurrent layers to reduce sequential computation and capture longer-range dependency. Later, Transformer-XL \cite{transformer-xl} is proposed by introducing the notion of recurrence into the deep self-attention network. Thus it enables the Transformer networks to capture the very long-term dependency during language modeling. 

\subsection{Multi-task Learning}
Multi-task learning is an approach for knowledge transfer across related tasks. It improves generalization by leveraging the domain-specific information contained in the training signals of related tasks \cite{caruana1997multitask}. Through sharing hidden layers among tasks, the model can capture the common features among all the tasks. Furthermore, by preferring the representation that all tasks prefer, the risk of over-fitting is reduced, and the model can be more general to new tasks in the future. Multi-task learning has been successfully used in many fields including natural language processing \cite{liu2015representation,guo2018soft,Devlin2019Bert}, speech recognition \cite{deng2013new} and computer vision \cite{long2015learning,lu2017fully}.

\begin{figure*}[t] 
\setlength{\abovecaptionskip}{0cm}
\centering\includegraphics[width=15.5cm]{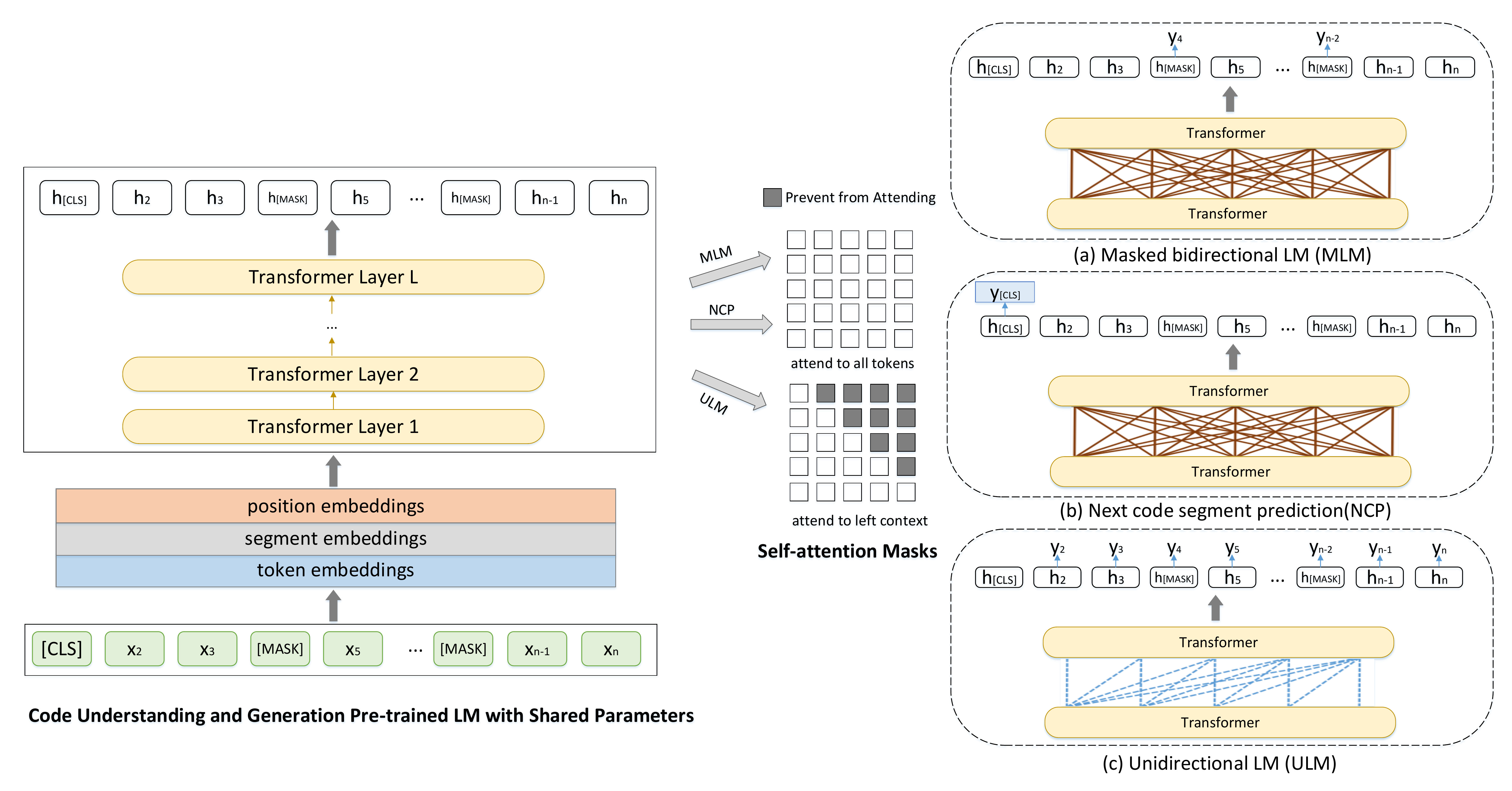} 
\caption{Overview of CugLM pre-training. The model parameters are shared across the pre-training objectives (i.e., MLM, NCP, and ULM). We use different self-attention masks to control the access to context for each token.}
\label{Fig:model}
\vspace{-0.3cm}
\end{figure*}

\subsection{Pre-trained Language Models}
Language model pre-training has shown to be effective for NLP, and has achieved the state-of-the-art results across many NLP tasks \cite{dai2015semi,Devlin2019Bert,Peters2018Deep,radford2018improving,Howard2018Universal}. The advantages of the pre-trained model can be summarized as follows: (1 By pre-training on the huge corpus, the model can learn universal representations and help with the target tasks; 2) The pre-trained model can provide a better model initialization, which leads to a better generalization performance on the downstream tasks. 3) Pre-training can be regarded as a kind of regularization to avoid over-fitting on small data. To apply the pre-trained language representations to downstream tasks, the feature-based approaches use the pre-trained representations as additional features \cite{Peters2018Deep}, and the fine-tuning approaches directly adapt the model on the downstream tasks by simply fine-tuning the pre-trained parameters \cite{radford2018improving,Devlin2019Bert}. Generative Pre-trained Transformer (GPT) \cite{radford2018improving} and Bidirectional Encoder Representations from Transformers (BERT) \cite{Devlin2019Bert} are the widely used fine-tuning approach, where BERT has significantly improved the performance of a wide range of natural language understanding tasks. However, the bidirectionality nature of BERT makes it difficult to be applied to natural language generation tasks. To overcome this limitation, UNIfied pre-trained Language Model (UNILM) \cite{dong2019unified} that can be applied to both natural language understanding (NLU) and natural language generation (NLG) tasks was proposed. Inspired by these models, we build a pre-trained language model for code understanding and generation, and then fine-tune it on code completion.

\section{\texorpdfstring{C\MakeLowercase{ug}LM}{CugLM}}\label{approach}
We describe the details about our proposed \textbf{C}ode \textbf{u}nderstanding and \textbf{g}eneration pre-trained \textbf{L}anguage \textbf{M}odel (\textbf{CugLM}) in this section.

\subsection{Model Architecture}
Given an input program token sequences $x=x_1,x_2,...,x_n$, CugLM obtains a contextualized vector representation for each token. The model architecture is shown in Figure \ref{Fig:model}. We adopt an $L$-layer Transformer as the language model to encode the input vectors $x = x_1, x_2,..., x_n$ into contextual representations at different levels $H^l = [h_1^l, h_2^l, ..., h_n^l]$, where $H^l=\text{Transformer}_l(H^{l-1}), l \in [1,L]$. In Figure \ref{Fig:model} and later sections, we omit the superscript $L$ for the hidden vectors of the final layer $h_i^L$ to make the illustration less cluttered. For each transformer layer (block), multi-attention heads are used to aggregate the output of the previous layer, and the output of a self-attention head $A_l$ is computed as: \\
\begin{equation}
\begin{split}
& Q = H^{L-1}W_l^Q,\ K= H^{L-1}W_l^K, \ V= H^{L-1}W_l^V \\
& M_{ij}=\left\{
    \begin{aligned}
    0, \ \text{allow to attend} \\
    -\infty, \text{prevent from attending}
    \end{aligned}
    \right. \\
& A_l = \text{softmax}(\frac{QK^T}{\sqrt{d_k}} + M)V
\end{split}
\end{equation}
where $H^i \in \mathbb{R}^{|x| \times d_h}$ denotes the $i$-th layer's output. The queries $Q$, keys $K$, and values $V$ are computed by linearly projecting the previous layer's output $H^{l-1}$ using parameter matrices $W_l^Q, W_l^K, W_l^V$. $M \in \mathbb{R}^{|x|\times|x|}$ is the mask matrix that determines whether a pair of tokens can be attended to each other. For different pre-training objectives, we use different mask matrices $M$ to control how many contextual tokens can a token attend to when computing its contextualized representations, as illustrated in Figure \ref{Fig:model}. For bidirectional LM, the elements of the mask matrix are all 0s, which means that all the tokens have access to each other. For unidirectional LM, the upper triangular part of the mask is set to $- \infty$, indicating that each token can only access the leftward context tokens and itself. The output of CugLM includes (1) contextual vector representation of each input token, and (2) the representation of \textsc{[CLS]}, which is short for ``CLaSsification'' and works as the aggregated sequence representation and can be used for classification tasks.

During the pre-training period, the model's parameters are shared and optimized with several objectives, namely, Masked bidirectional LM, Next Code segment Prediction, and Unidirectional LM. After the model is pre-trained, we can then fine-tune it for downstream tasks. In this paper, we fine-tune CugLM on code completion.

\begin{figure*}[t] 
\setlength{\abovecaptionskip}{0cm}
\centering\includegraphics[width=15cm]{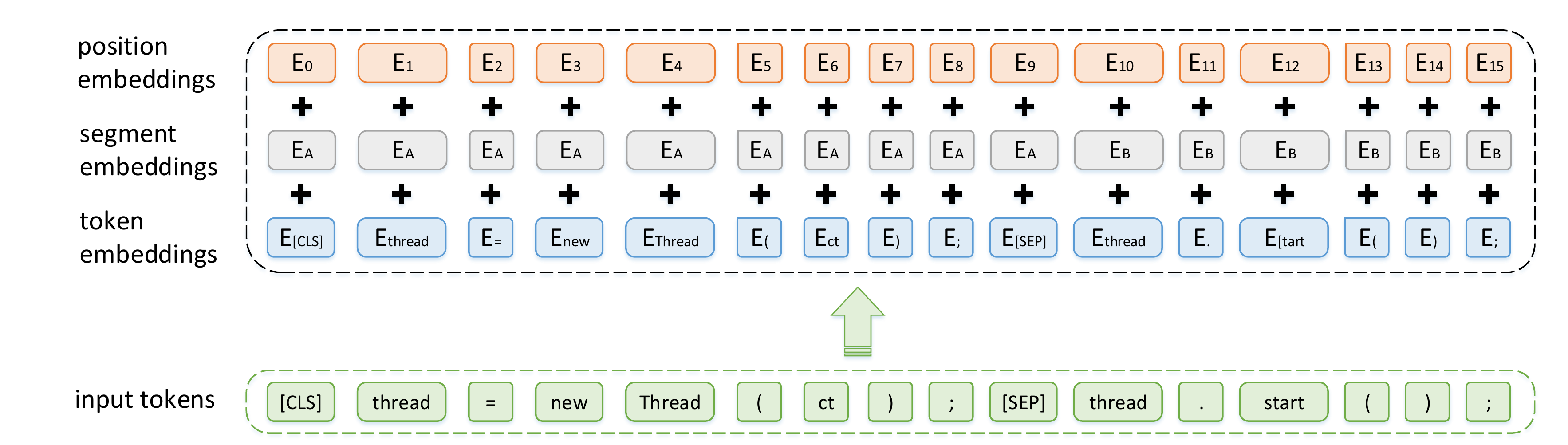} 
\caption{Input representation. The input embeddings is the sum of the token embeddings, the segment embeddings, and the position embeddings.}
\label{Fig:input}
\vspace{-0.3cm}
\end{figure*}

\subsection{Input Representation}
The input $x$ is a token sequence, which is a pair of segments packed together. As shown in Figure \ref{Fig:input}, for a given token, its vector representation is computed by summing the corresponding token, segment and position embeddings.  

\begin{itemize}
    \item For token embeddings, the embedding matrix is randomly initialized, and then is adjusted as part of the training process. Two special tokens \textsc{[CLS], [SEP]} are defined, where \textsc{[CLS]}, which is short for ``CLaSsification'', always appears at the beginning of the input. The final hidden state corresponding to \textsc{[CLS]} can be used as the aggregate sequence representation for classification tasks, for example, in next sentence prediction task. \textsc{[SEP]}, which is short for ``SEPeration'', is used to separate the sentence pairs.
    \item The segment embeddings, i.e., $E_A$ and $E_B$ are also used to differentiate the code segment pairs. For each token of the first code segment, a learned embedding $E_A$ is added, and a learned embedding $E_B$ is added to each token of the second code segment. The embedding matrix for the segment embeddings is also randomly initialized 
    \item To make use of the order of the sequence, we use learned positional embeddings with sequence lengths up to 128 tokens.
\end{itemize}

\subsection{Pre-training Procedure}
To pre-train CugLM, we adopt multi-task learning to learn three tasks jointly, as shown in Figure \ref{Fig:model}, including Masked bidirectional Language Modeling (MLM), Next Code segment Predicting (NCP), and Unidirectional Language Modeling (ULM). For the first two objectives, the Transformer network is under the bidirectional settings, and for the last objective, the Transformer network is unidirectional.

\noindent a) \textbf{Masked bidirectional Language Modeling}: In order to train deep bidirectional representations for the program, we adopt a similar objective with BERT, that is, masking some percentage of the input tokens and then predicting only those masked tokens. Different from BERT, we only mask the identifiers with type information, where the type information can be extracted by static analysis or be annotated by developers, considering that these identifiers are more informative for understanding the program. Then the objective is to predict the masked identifiers based on their bidirectional contextual tokens, where all tokens can attend to each other in prediction. It encodes contextual information from both directions and can generate better contextual representations of the masked identifiers as well as the other tokens than its unidirectional counterpart. The final hidden vectors corresponding to the mask tokens are fed into the output \textit{softmax} layer to produce the probability distribution of the outputs.

\noindent b) \textbf{Next Code segment Predicting}: Understanding the relationship between two sentences is quite important for many NLP tasks, for example, Question Answering (QA) and Natural Language Inference (NLI), which can help to understand the input text in more depth. We argue that understanding relationships between code segments also help in source code modeling. In order to achieve this, we pre-train a binarized next code segment prediction task, that is, predicting whether two segments of code tokens follow each other in a piece of code snippet. Specifically, when choosing the code segments A and B for each pre-training example, 50\% of the time B is the actual next code segment that follows A, and 50\% of the time it is a random code segment from the corpus. For example: \\

\noindent \textbf{Input} = [CLS] public void setTextDirection ( int textDirection ) \{
\hspace*{1cm} [SEP] this . mTextDirection = textDirection ; \}
        
\noindent \textbf{Label} = 1 \\

\noindent \textbf{Input} = [CLS] public void setTextDirection ( int textDirection ) \{ 
\hspace*{1cm} [SEP] this . request = request ; 
        
\noindent \textbf{Label} = 0 \\

The final hidden vector corresponding to \textsc{[CLS]}, which works as the aggregated sequence representation, is fed into the output \textit{softmax} layer to produce the probability distribution of classification results.

\noindent c) \textbf{Unidirectional Language Modeling}: For language generation tasks, for example, code completion, the context of the predicted token should only consist of the token on its left. Thus, we create the left-to-right language modeling task as another pre-training objective, namely predicting the next token $x_{t+1}$ given the preceding context tokens ${x_1, x_2, ..., x_{t}}$. The representation of each token encodes only the leftward context tokens and itself. This can be done using a triangular matrix for self-attention mask $M$, where the upper triangular part of the self-attention mask is set to $-\infty$, and others to 0. At each time step $t$, the final hidden vector corresponding to $x_t$ is fed into the \textit{softmax} layer to produce the probability distribution of the predicted token $y_t$. 

The pre-training procedure follows the existing language model pre-training approaches. The parameters of CugLM are learned to minimize the sum of the cross-entropy losses of the three pre-training tasks, and are shared among all the tasks. The final loss function is given below:

\begin{equation}
\min_\theta \mathcal{L}_{MLM}(\theta) + \mathcal{L}_{NCP}(\theta) + \mathcal{L}_{ULM}(\theta) 
\end{equation}

\subsection{Fine-tuning Procedure}
\begin{figure}[h] 
\setlength{\abovecaptionskip}{0cm}
\centering\includegraphics[width=7.5cm]{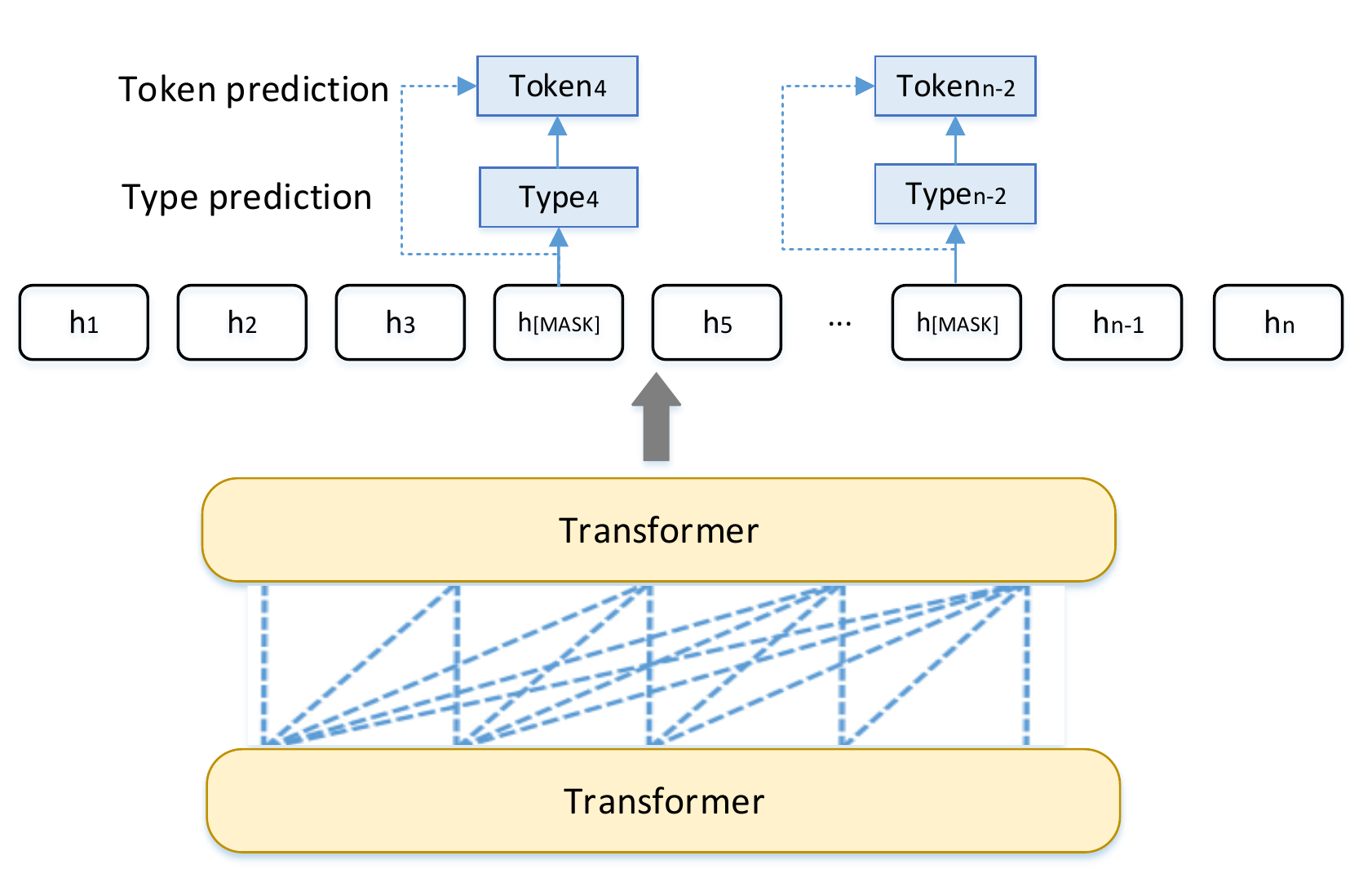} 
\caption{Model architecture for UMLM.}
\label{Fig:UMLM}
\vspace{-0.3cm}
\end{figure}

When the model is pre-trained, we fine-tune it on code completion task. In code completion, the context of the predicted token should only consist of all the token on its left. Thus, the representation of each token can encode only the leftward context tokens and itself. During the fine-tuning procedure, the following two objectives are optimized:

\noindent a) \textbf{Unidirectional Masked Language Modeling (UMLM)}: Different from the MLM objective in pre-training, the UMLM objective in fine-tuning is to predict the masked token based only on its leftward context, where all tokens can only attend to the tokens on its left in prediction. The transformer network is set to unidirectional using a triangular matrix for the self-attention mask. All the identifiers that have type information are masked in each sequence. Besides, our model not directly predicts the masked token. Instead, we adopt the multi-task learning framework to predict the token and its type. We first predict the type of the token, and then the predicted type is used to assist the token prediction, as shown in Figure \ref{Fig:UMLM}. The reason for formulating the code completion task as a two-step prediction instead of predicting the type and token jointly lies in that, by predicting the type firstly and then use the predicted results as extra input for the token prediction can constraint our model to make more accurate prediction on the type and further enhance the token prediction performance.

\noindent 1) Type prediction: The final hidden vector (i.e., the output of the Transformer) corresponding to the mask token $h_{[MASK]}$ is used to compute the output vector for the token's type $O_{type}$. We use the \textit{softmax} function to produce the probability distribution of the outputs $Y_{type}$:  
\begin{equation}\label{formula:y}
\begin{split}
O_{type} & = \tanh(W^o h_{[MASK]}) \\
Y_{type} & = \text{softmax}(W^y O_{type} +b^y) 
\end{split}
\end{equation}
where $W^o\in\mathbb{R}^{H \times H_{type}}, W^y\in\mathbb{R}^{V_{type} \times H_{type}},  b^y\in\mathbb{R}^{V_{type}} $ are trainable parameters. $V_{type}$ is the vocabulary size of the token's type, $H$ is the hidden size of the transformer network, $H_{type}$ is the embedding size of type vector.

\noindent 2) Token prediction: After predicting the token's type, we use the predicted type to assist the token prediction. The vector of the predicted type $E_{type}$ and the hidden vector of the mask token $h_{[MASK]}$ are concatenated to compute the output vector for the token $O_{token}$. Then the output vector is fed into the output \textit{softmax} layer to compute the output vector for the token $Y_{token}$:

\begin{equation}\label{formula:y}
\begin{split}
O_{token} & = \tanh(W^o (h_{[MASK]};E_{type}) ) \\
Y_{token} & = \text{softmax}(W^y O_{token} +b^y) 
\end{split}
\end{equation}
where $E_{type}$ is the embedding of the predicted type, 
$W^o\in\mathbb{R}^{H_{token} \times H}$, $W^y\in\mathbb{R}^{V_{token} \times H_{token}}$,  $b^y\in\mathbb{R}^{V_{token}}$ are trainable parameters. $V_{token}$ is the vocabulary size of the token, and ``;'' denotes the concatenation operation. 

\noindent b) \textbf{Unidirectional Language Modeling (ULM)}: This objective is a left-to-right language modeling task that is the same as the pre-training procedure. Given the preceding context tokens $x_1, x_2, ..., x_t$, the model predicts the next token $x_{t+1}$, where the representation of each token encodes only the leftward context tokens and itself.

During the fine-tuning procedure, the parameters of CugLM are learned to minimize the sum of the cross-entropy losses of the two fine-tuning tasks and are shared among all the tasks. The final loss function is given below:

\begin{equation}
\min_\theta \mathcal{L}_{UMLM}(\theta) + \mathcal{L}_{ULM}(\theta)
\end{equation}

Through learning these two objectives jointly, we hope the model can make better predictions on both the identifiers and the other tokens.

\section{Experiments and analysis}\label{exp}

\subsection{Data preparation}
We pre-train and fine-tune our model across two programming languages: Java and TypeScript. The programs in the corpus are collected from publicly available open-source GitHub repositories by removing duplicate files and project forks. Each program is tokenized into token sequence. The detailed information is shown in Table \ref{tab:datasets}. We use 60\% of the projects for pre-training, and 40\% of the projects for fine-tuning on code completion task. During the fine-tuning, we split the projects into train/validation/test sets in the proportion 8:1:1. For the other baselines, all the programs used in pre-training and the training programs used in fine-tuning are used as the training set, and the validation and test sets are the same as in our fine-tuning procedure. We also randomly sample 200 program files from both Java and TypeScript test sets as the small test sets for Byte Pair Encoding based Neural Language Model (BPE NLM) \cite{karampatsis2020big} evaluation since when performing completion (testing) in their model, they use a variation of the beam search algorithm to combine the sub-units to complete tokens, which is very time-consuming. It takes several minutes to complete a single program file and will take tens of days to perform completion on the large test sets (e.g., the Java test set contains 14,600 files). Thus, we create small test sets.

\begin{table}[t]
\centering
\setlength{\abovecaptionskip}{0cm} 
\caption{Statistics of the datasets.}
\begin{tabular}{lcc}  
\toprule
 ~ & Java & TypeScript \\
\midrule
Projects & 9,708 & 8,446 \\
Files & 800,983 & 227,424\\
Lines & 5.4 * $10^7$ & 8.8 * $10^6$  \\
\# of Tokens & 6.9 * $10^6$ & 1.1 * $10^6$ \\
\# of Types &  6.4 * $10^6$ & 1.7 * $10^5$ \\
Masked ID proportion & 21.04\% & 9.74\% \\
\bottomrule
\end{tabular}
\vspace{-0.3cm}
\label{tab:datasets}
\end{table}

\begin{figure}[t] 
\setlength{\abovecaptionskip}{0cm}
\centering\includegraphics[width=7.3cm]{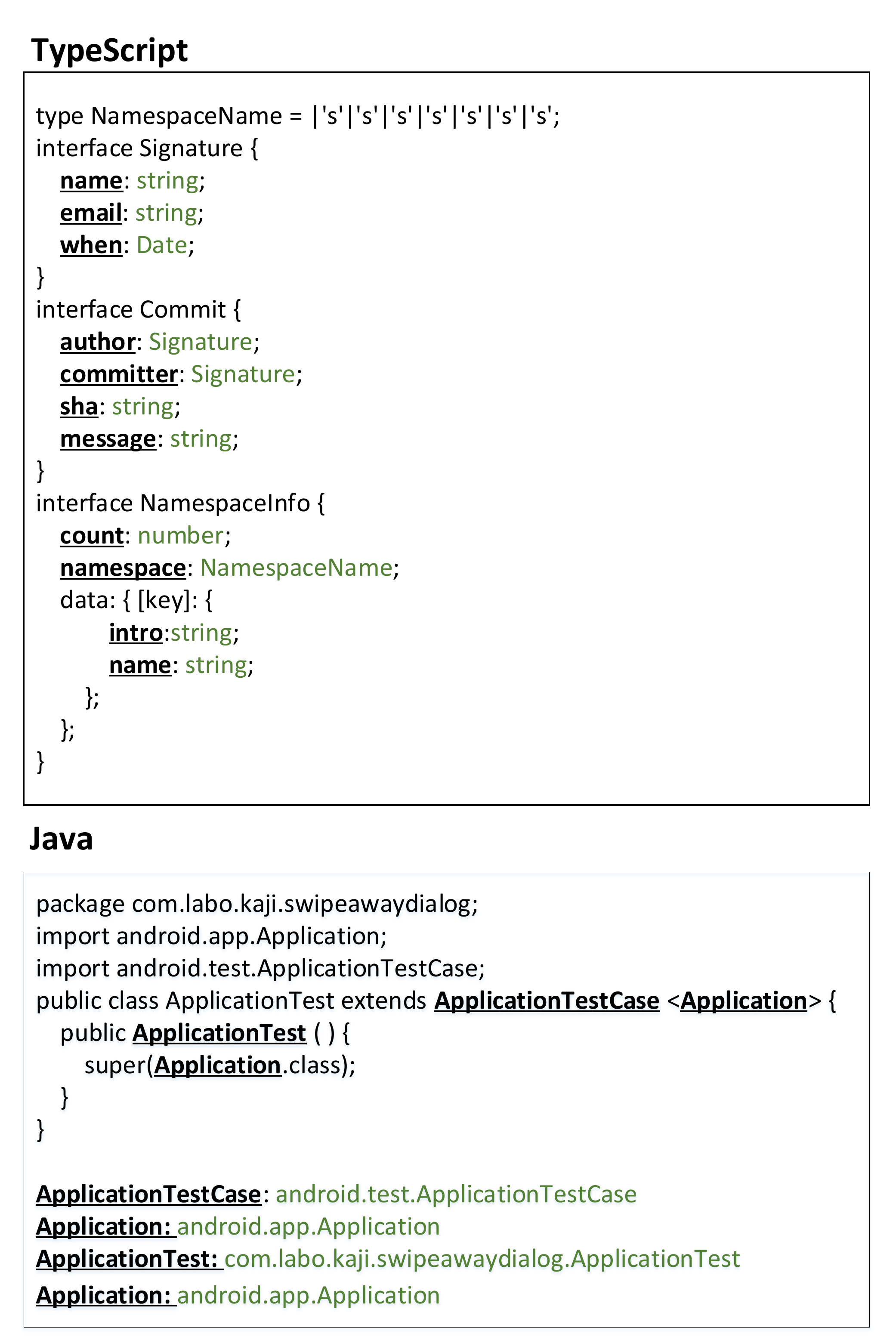} \caption{Code examples for type annotations.}
\label{Fig:code_ep}
\vspace{-0.3cm}
\end{figure}

\begin{table*}[t]
  \begin{center}
  \setlength{\abovecaptionskip}{0cm} 
  \caption{Performance of baseline models and our approach.}
    \begin{tabular}{lcccccccc} 
      \toprule
      \multirow{3}{*}{Model} & \multicolumn{4}{c}{\textbf{Java}} & \multicolumn{4}{c}{\textbf{TypeScript}} \\
      \cmidrule(lr){2-5} \cmidrule(lr){6-9}
      ~ & \multicolumn{2}{c}{\textbf{Large Test}} & \multicolumn{2}{c}{\textbf{Small Test}} & \multicolumn{2}{c}{\textbf{Large Test}} & \multicolumn{2}{c}{\textbf{Small Test}} \\
      \cmidrule(lr){2-5} \cmidrule(lr){6-9}
      ~ & All Tokens & Identifiers & All Tokens & Identifiers & All Tokens & Identifiers & All Tokens & Identifiers \\
      \midrule
      Vanilla LSTM & 64.28\% & 33.84\% & 64.73\% & 33.11\%  & 64.42\% & 28.11\% & 63.31\% & 23.91\% \\
      Pointer Mixture Network & 68.30\% & 38.41\% & 68.49\% & 37.54\% & 68.75\% & 33.76\% & 65.75\% & 29.26\% \\
      BPE NLM & - & - & 67.17\% & 43.67\% & - & - &  65.39\% & 36.16\% \\
      Transformer-XL & 72.12\% & 43.63\% & 70.96\% & 40.92\% & 73.94\% & 37.46\% & 68.88\% & 34.90\% \\ 
      \midrule
      CugLM & 84.06\% & 55.19\% & 80.07\% & 48.47\% & 82.14\% & 41.85\% & 81.36\% & 39.28\% \\
      \bottomrule
    \end{tabular}
    \label{tab:results}
    \vspace{-0.3cm}
  \end{center}
\end{table*}

For Java programs, we extract the identifiers' type information through static analysis. For TypeScript programs, we apply the approach in \citet{hellendoorn2018deep} to extract type annotations of the identifiers. We filter the programs to make sure at least 10\% of type annotations are user-defined types in each TypeScript file. Figure \ref{Fig:code_ep} shows the examples for Java and TypeScript code, where the identifiers that have type are marked with underlines, and the green tokens next to the identifiers are the corresponding types. To generate each training input sequence for pre-training, we sample two spans of tokens from the corpus, which we refer to as segments $S_1$ and $S_2$. Each segment contains several lines of source code tokens. For the first segment $S_1$, we sample the first $N$ lines from one program file, where $N$ is randomly sampled from 1 to the length of the code lines of the program file. 50\% of the time, the second segment $S_2$ is the rest of the lines from the same program file that follows $S_1$, and 50\% of the time it is a random code segment sampled from other program files of the corpus, which is done for the ``next code segment prediction (NCP)'' task. They are sampled such that the combined length is $\leq$ 128 tokens. For the ``Masked bidirectional Language Modeling (MLM)'' task, we only mask those identifiers that have type information. For example, the underlined tokens in Figure \ref{Fig:code_ep}. 

\subsection{Experimental Setup}
\noindent\textbf{Parameter configuration.} We use Transformer with 6 layers, 516 dimensional hidden states and 6 attention heads. The inner hidden size of the feed-forward layer is 3072. We pre-train our model with batch size of 16 sequences for 600,000 steps. We use Adam with learning rate of 5e-5, $\beta _1$ = 0.9, $\beta _2$ = 0.999, L2 weight decay of 0.01, learning rate warmup over the first 1,000 steps, and linear decay of the learning rate. We use a dropout probability of 0.1 on all layers. We use a gelu activation \cite{hendrycks2016bridging} following OpenAI GPT. The training loss is the sum of the cross-entropy losses of the pre-training objectives or fine-tuning objectives. Training of CugLM was performed on 3 GeForce GTX 1080 Ti GPUs with 12GB memory. For each dataset, the model is pre-trained for 600,000 steps and takes 4 days to complete, and is fine-tuned for 300,000 steps and takes 2 days to complete.

\noindent\textbf{Metric.} We use \textit{accuracy} to evaluate the performance of code completion. Our model provides an ordered list of suggestions for each token in the source code file given the context. We compute the top-1 accuracy, i.e., the fraction of times the correct suggestion appears in the first of the predicted list. 

\noindent\textbf{Vocabulary.} ~ As shown in Table \ref{tab:datasets}, in the datasets, the number of unique tokens and types is too large to build neural models to learn directly. We choose $K$ (50,000) most frequent tokens in each training set to build the token vocabulary, which is the same as \citet{Li2018Code}'s study. For those tokens outside the vocabulary, we use \emph{UNK} (unknow values) to represent them. The size of type vocabulary is also set to 50,000. In both the training and test process, the predictions of the \textit{UNK} targets are treated as wrong predictions. The token \emph{UNK} rates for Java, and TypeScript test sets are 10\%, 5\%, and the type \emph{UNK} rates are 9\%, 1\%, respectively.

\subsection{Research Questions and Results}
To evaluate our proposed approach, in this section, we conduct experiments to investigate the following research questions:

\noindent\textbf{RQ1: How does our proposed approach perform in code completion when compared with state-of-the-art models?} ~ To answer this research question, we compare our model with the following baseline models: 
\begin{itemize}
    \item vanilla LSTM: a vanilla LSTM neural network-based language model.
    \item Pointer Mixture Network \cite{Li2018Code}: an attention and pointer-generator network-based code completion model.
    \item Byte Pair Encoding based Neural Language Model (BPE NLM) \cite{karampatsis2020big}: a large-scale open-vocabulary NLM for code completion, which leverage BPE \cite{gage1994new} algorithm to keep vocabulary size low and successfully predict OoV (Out-of-Vocabulary) tokens. 
    \item Transformer-XL \cite{transformer-xl}: a self-attentional neural network-based language model for code completion.
\end{itemize}

\begin{table*}[t]
  \begin{center}
  \setlength{\abovecaptionskip}{0cm} 
  \caption{Effects of each pre-training task, fine-tuning task, and the type prediction in our proposed model.}
    \begin{tabular}{lcccccccc} 
      \toprule
      \multirow{3}{*}{Model} & \multicolumn{4}{c}{\textbf{Java}} & \multicolumn{4}{c}{\textbf{TypeScript}} \\
      \cmidrule(lr){2-5} \cmidrule(lr){6-9}
      ~ & \multicolumn{2}{c}{\textbf{Large Test}} & \multicolumn{2}{c}{\textbf{Small Test}} & \multicolumn{2}{c}{\textbf{Large Test}} & \multicolumn{2}{c}{\textbf{Small Test}} \\
      \cmidrule(lr){2-5} \cmidrule(lr){6-9}
      ~ & All Tokens & Identifiers & All Tokens & Identifiers & All Tokens & Identifiers & All Tokens & Identifiers \\
      \midrule
      Full Model & 84.06\% & 55.19\% & 80.07\% & 48.47\% & 82.14\% & 41.85\% & 81.36\% & 39.28\% \\
      \midrule
      Pre-training tasks  \\
      - ULM & 78.64\% & 50.10\% & 77.78\% & 44.18\% & 77.83\% & 38.44\% & 76.77\% & 37.38\% \\
      - MLM & 77.42\% &	49.86\% & 76.41\% & 43.82\% & 78.93\% & 36.89\% & 78.28\% & 35.15\% \\
      - NCP & 81.24\% & 52.56\% & 79.79\% & 46.54\% & 78.52\% & 40.71\% & 79.02\% & 38.49\% \\
      \midrule
      Fine-tuning tasks \\
      - UMLM & 80.93\% & 45.70\% & 	77.21\%	& 41.66\% & 78.86\% & 33.26\% &	77.58\% & 31.81\% \\
      - ULM & - & 49.50\% & - & 43.31\% & - & 38.25\% & - & 35.33\% \\
      \midrule
      - Type Prediction & 80.14\% & 52.05\% & 77.28\% & 46.83\% & 80.99\% & 40.73\% & 79.85\% & 38.31\% \\
      \bottomrule
    \end{tabular}
    \label{tab:ablation_results}
    \vspace{-0.3cm}
  \end{center}
\end{table*}

\textit{ 1) Comparison with LSTM based closed vocabulary models (the first two baselines): } To compare with Pointer Mixture Network, we downloaded their publicly available source code\footnote{\text{https://github.com/jack57lee/neuralCodeCompletion}}. In their model, the programs in the datasets are parsed into ASTs, and they build the model to perform code completion on AST node sequences. Although the ASTs can provide more information, representing the programs as AST node sequences is not the natural order of typing, and the precision does not directly reflect the productivity gain of the code completion tool. More importantly, in practice, the code is incomplete, so the software project might not be compilable (code is not parsable into ASTs, or parsed ASTs miss a lot of information). Thus, representing programs as token sequences and performing code completion on the token-level might be more practical. In this paper, we focus on token-level code completion. In our corpus, the programs are tokenized into token sequences. To compare with them, we train their model within our tokenized programs using the command line arguments given in the artifact’s README file \footnote{Since the Pointer Mixture Network also makes use of the additional information derived from ASTs, the results of using token sequence as input might understate the accuracy of the plain Pointer Mixture Network.}. Their base model is a single layer LSTM network with an unrolling length of 50 and hidden unit size of 1500. The initial learning rate is 0.001 and is decayed by multiplying 0.6 after every epoch. The gradients’ norm is clipped to 5. The size of the attention window is 50. Since the Pointer Mixture Network is based on LSTM language model, we also list the results of the vanilla LSTM, where the parameter configuration of the vanilla LSTM network is set the same as the Pointer Mixture Network. 

As shown from the results, our model outperforms the two LSTM-based models on both Java and TypeScript datasets by a large margin, especially in identifier completion. On the Java large test set, our model achieves the accuracy of 84.06\% and 55.19\% on token's completion and identifier's completion, respectively, which outperforms Pointer Mixture Network by 23.07\% and 43.69\%, in terms of relative improvement. On the TypeScript large test set, our model achieves the accuracy of 82.14\% and 41.85\% on token's completion and identifier's completion, respectively, which outperforms Pointer Mixture Network by 19.47\% and 23.96\%. The results on small test sets are similar to the large test set. We can find that the improvements on the TypeScript dataset are smaller than Java, especially in identifier completion. The reason lies in that, the (masked) identifier proportion in TypeScript (9.74\%) is smaller than Java (21.04\%) because the type information in TypeScript is annotated by developers, and only a part of the identifiers are annotated. In the MLM pre-training task, these identifiers are masked and are predicted based on their contextual tokens aiming at generating better contextual and informative representations for these identifiers as well as other tokens. During fine-tuning, the type information of these identifiers is used to assist the identifiers' prediction. Due to the lower masked proportion, the pre-training and fine-tuning procedure can offer less information than Java, thus resulting in smaller improvements.

\textit{ 2) Comparison with open vocabulary model (BPE NLM): } 
To compare with BPE NLM, we downloaded their publicly available source code\footnote{https://github.com/mast-group/OpenVocabCodeNLM} and train their model on our datasets. They use a single layer GRU NLM with an unrolling length of 200 built upon sub-word units learned from BPE. The embedding size and the hidden unit size are both set to 512 in their model. To keep the number of parameters comparable with our model and other baselines, we increase the hidden unit size and the embedding size of their model to 1500. There are three scenarios: static, dynamic, and maintenance, where the dynamic and maintenance settings update model's parameters during testing. Since our model and other baselines do not update parameters during the test process, we present the results of the static scenario to make the comparison fair, and realize that evaluating dynamically may improve accuracy. As shown from the results, BPE NLM performs best on completing identifiers among all the baseline models on both datasets, which proves the power of the open vocabulary LM for predicting the identifiers. Even though, our model still outperforms the BPE NLM on completing identifiers. When evaluating on completing all kinds of tokens, the performance of BPE NLM is not as well as the identifier completion. Our model outperforms BPE NLM on completing all kinds of tokens by a large margin.

\textit{ 3) Comparison with transformer network based model (Transformer-XL): } To find out if CugLM’s promising results derive more from using a Transformer-based model for code completion, or from the multi-task learning based pre-training and fine-tuning, we compare our results to a Transformer-based model trained from scratch, i.e., without the benefit of a pre-trained embedding. Transformer-XL is a Transformer network based language model, which introduces the notion of recurrence to model the long-term dependency of the input sequence. We use a 6-layer Transformer-XL network with 5 parallel heads. The dimension of each head is set to 64. We set the segment length to be 128, and the length of the cached segments to 256. The dimensionality of the model (hidden unit) and the embedding size is set to 800. The dimension of the feed-forward layer is set to 1024. As seen from Table \ref{tab:results}, transformer-XL model outperforms the other baseline models that are based on the recurrent neural networks on both datasets, which demonstrates that the Transformer-based network is more powerful than recurrent neural networks on this task. The performance of our model is substantially higher than the Transformer-XL model trained from scratch. We therefore conclude that pre-training and fine-tuning are crucial to CugLM’s success.

\noindent\textbf{RQ2: What are the contributions of the pre-training tasks? } ~ We perform an ablation study to examine the effects of the three pre-trained tasks: ULM, MLM, and NCP. We conduct experiments on pre-training the model without each task, and the fine-tuning procedure remains unchanged. The results are shown in Table \ref{tab:ablation_results}. The first row shows the results of our full model. The second to the fourth rows present the results of removing ULM, MLM, and NCP from the full model during pre-training, respectively. 

\noindent\textbf{- ULM} ~ Removing the ULM task during pre-training. The loss function of the pre-training procedure consists of $\mathcal{L}_{MLM}$ and $\mathcal{L}_{NCP}$, and both these tasks are based on the bidirectional transformer. As seen from the results, removing this task hurts the model's performance. During fine-tuning, the objectives are based on the unidirectional transformer. Thus, adding the ULM task during pre-training makes the learned text representations more general because they are optimized for both bidirectional and unidirectional language modeling objectives jointly, mitigating over-fitting to bidirectional language modeling task. Removing the ULM task would make the parameters hard to optimized when fine-tuned on the unidirectional objectives. Thus, the accuracy drops. 

\begin{table*}[t]
  \begin{center}
  \setlength{\abovecaptionskip}{0cm} 
  \caption{Performance of completing different types of tokens.}
    \begin{tabular}{lcccccc} 
     \toprule
     ~ & \multicolumn{3}{c}{\textbf{Java}} & \multicolumn{3}{c}{\textbf{TypeScript}} \\
     \cmidrule(lr){2-4} \cmidrule(lr){5-7}
      Type & Proportion & CugLM & BPE NLM & Proportion & CugLM & BPE NLM \\
      \midrule
      Identifiers & 28.99\% & \textbf{48.47\%} & 42.27\% & 16.62\% & \textbf{39.28\%} & 36.16\% \\
      Keyword & 7.69\% & \textbf{86.78\%} & 72.57\% & 6.49\% & \textbf{79.47\%} & 56.67\% \\
      Punctuation & 31.98\% & 87.38\% & \textbf{90.30\%}  & 45.42\% & 82.64\% & \textbf{82.93\%} \\
      Numerals & 0.62\% & \textbf{72.62\%} & 58.83\% & 1.22\% & \textbf{89.42\%} & 82.44\%\\
      Operator & 3.80\% & \textbf{85.65\%} & 76.92\% & 4.03\% & \textbf{75.98\%} & 65.84\% \\
      \bottomrule
    \end{tabular}
    \label{tab:completion_types}
    \vspace{-0.3cm}
  \end{center}
\end{table*}

\noindent\textbf{- NCP} ~ Removing the Next Code segment Prediction task during the pre-training. The loss function consists of $\mathcal{L}_{ULM}$ and $\mathcal{L}_{MLM}$. The NCP tasks are added to help the model understand the relationships between the code segments. The model removing NCP performs worse than the full model, but performs better than removing ULM, which demonstrates that the NCP task is necessary to improve the performance but contributes less than the ULM task.

\noindent\textbf{- MLM} ~ Pre-training the model without the Masked bidirectional Language Modeling objective, and the loss function consists of $\mathcal{L}_{ULM}$ and $\mathcal{L}_{NCP}$. As shown from the results, removing the MLM hurts the performance more than the other two tasks, especially on identifier completion. MLM task can help the model generate better contextual representations of the tokens, especially the identifiers, thus can improve the model's performance significantly.

The above results demonstrate that all of the pre-training tasks are necessary to improve the performance, and MLM contributes most to the improvements. 

\noindent\textbf{RQ3: What are the contributions of the fine-tuning tasks? } To figure out the effectiveness of the fine-tuning procedure, we also conduct experiments by removing each of the fine-tuning task. The results are shown in fifth and sixth rows of Table \ref{tab:ablation_results}.

\noindent\textbf{- UMLM} ~ Removing the Unidirectional Masked Language Modeling task during fine-tuning procedure. Only the left-to-right language modeling task is performed and the loss function becomes $\mathcal{L}_{ULM}$. As seen from the results, removing this task hurts the model's performance on both two datasets, especially for the identifier prediction. UMLM task can help the model generate better contextual representations for the tokens. Besides, it can also utilize the type information of the identifiers during the fine-tuning. Thus, this fine-tuning task is necessary for improving the performance of the code completion.

\noindent\textbf{- ULM} ~ Removing the Unidirectional Language Modeling task during fine-tuning procedure. Under this setting, the model can only produce the results of the masked identifier prediction. The loss function becomes $\mathcal{L}_{UMLM}$. As seen from the results, when removing ULM task, the performance of the identifier prediction drops a lot, which demonstrates that the language modeling task can offer much help for the identifier prediction. Through optimizing the model on this task jointly, the model can capture the semantic of the input code segment better, which serves as the basis of the improvement on identifier prediction.

\noindent\textbf{RQ4: Could the predicted type help the model on token prediction?} ~ When fine-tuning our model on code completion task, we utilize multi-task learning to predict the token and its type jointly. We first predict the type and then use the type to assist the token's prediction. To confirm whether our model can correctly predict the identifier's type, we present the accuracy of the type prediction. Our model achieves the accuracy of 68.89\% and 79.31\% on Java and TypeScript large test sets, respectively. The results demonstrate that our model can correctly predict the identifiers' type in most cases. To find out whether the type prediction really helps, we conduct experiment by removing the type prediction. The results are shown in the last row of Table \ref{tab:ablation_results}. As shown from the results, when removing the type prediction, the model performs worse than the full model on completing both identifiers and all tokens, which demonstrates that the predicted type information can help the model achieve better performance on code completion.

\section{Discussion}\label{discussion}

\subsection{The type of completions}
Except for identifiers, we also give a detailed breakdown of the accuracies for completing different types of tokens on both our model and BPE NLM \cite{karampatsis2020big}, and also present these tokens' proportion. The results are shown in Table \ref{tab:completion_types}. Punctuations make up the majority of the completions, and the accuracies of both our model and BPE NLM on predicting punctuations are high, where BPE NLM performs better than CugLM. The punctuation tokens are much easier to complete than identifiers, but these completions are not that useful for developers \cite{karampatsis2020big}. For keyword completions, our model outperforms BPE LM by a large margin. The keywords are predefined, reserved words used in programming that have special meanings to the compiler, which contain the syntactic information or the attribute information of the objects. The great performance of CugLM on completing keywords further demonstrates that through multi-task learning based pre-training and fine-tuning, the representations generated by our model can capture syntactic and semantic information better. For numeral and operator completions, which are more related to the semantic of the programs, our model also outperforms BPE NLM substantially.

\subsection{Model complexity comparison}
To analyze the complexity of our model and the baseline models, we present the number of trainable parameters for all the models, shown in Table \ref{tab:param}. The number of trainable parameters of our model is less than all the baselines. Although we adopt multi-task learning for both pre-training and fine-tuning, the number of trainable parameters does not increase much as all of the tasks share one multi-layer transformer network. To improve training efficiency and avoid over-fitting, we do not use large parameter settings. Even though, our model still outperforms the other baselines by a large margin thanks to the pre-training and fine-tuning.

\begin{table}[h]
  \begin{center}
  \setlength{\abovecaptionskip}{0cm} 
  \caption{Parameters of the baseline models and our model.}
    \begin{tabular}{lcc} 
     \toprule
      Model & \# of Parameters \\
      \midrule
      Vanilla LSTM & 168M \\
      Pointer Mixture Network & 177M\\
      BPE NLM & 145M \\
      Transformer-XL & 173M \\
      \midrule
      CugLM & 104M\\
      \bottomrule
    \end{tabular}
    \label{tab:param}
    \vspace{-0.3cm}
  \end{center}
\end{table}

\subsection{Effect of applying BPE algorithm}
To further improve the performance of our model, we also conduct experiments on applying Byte Pair Encoding (BPE) algorithm to build up the vocabulary of sub-words as in \cite{karampatsis2020big}, where the rare tokens will be segmented into more common sub-word units, and no word is OoV. However, the performance on Java corpus is comparable with the origin model, and the accuracy decreases slightly on TypeScript corpus. We analyze the possible reasons are as follows. During pre-training, we mask the identifiers with type information. When we apply BPE algorithm, most of these masked identifiers will be split into sub-word units. Thus, all of these units will be masked, which leads to the high mask proportion and increased the difficulty of learning the semantics of embeddings. Besides, during fine-tuning, our model utilizes the predicted type information to assist the token's prediction. After splitting the tokens into sub-word units, all of the units from one token correspond to the same type, resulting in the semantic inconsistencies between the type information and the sub-word units. For example, the same unit from different tokens might correspond to different types. Thus, applying BPE does not improve the performance of our model.

\subsection{Threats to Validity}
\noindent\textbf{Threats to external validity} relate to the quality of the datasets we used and the generalizability of our results. We create two massive datasets (Java and TypeScript) to pre-train and fine-tune our model. All of the programs in the datasets are collected from GitHub repositories. The reasons for using these two languages are as follows. These two languages are commonly used for software development, and we can get the identifiers' type through static analysis or through the developers' annotations. However, further studies are needed to validate and generalize our findings to other programming languages.

\noindent\textbf{Threats to internal validity} include the influence of the hyper-parameters used in our model. The performance of our model would be affected by different hyper-parameter settings, which are tuned empirically in our experiments. Thus, there is little threat to the hyper-parameter choosing, and there might be room for further improvement. However, current settings have achieved a considerable performance increase. Another threat to internal validity relates to the implementation of the baseline methods. For \citet{Li2018Code}'s model, we apply their model to the token-level code completion, which is originally used for AST-level code completion. In their model, the additional information derived from ASTs is utilized to improve the performance. The results of using token sequence as input might understate the accuracy of the plain Pointer Mixture Network.
However, in practice, the code is incomplete, so the code is not parsable into ASTs, or parsed ASTs miss a lot of information. Thus, representing programs as token sequences and performing code completion on the token-level is more practical. Under this setting, we have tried our best to make fair comparison with \citet{Li2018Code} by only changing the format of the input, and keeping the model unchanged. For BPE NLM \cite{karampatsis2020big}, we compare our model with the static setting of their model considering the fairness of the comparison. We realize that evaluating dynamically may improve accuracy. The dynamic and maintenance scenarios are not implemented and compared in this work, which will be considered as our future work. 

\noindent\textbf{Threats to construct validity} relate to the suitability of our evaluation measure. We use \textit{accuracy} as the metric which evaluates the proportion of correctly predicted next token. It is a classical evaluation measure for code completion and is used in almost all the previous code completion work \cite{hindle2012naturalness, tu2014localness, raychev2016probabilistic, hellendoorn2017deep, Li2018Code}. 

\section{Related Work}\label{related work}

\noindent\textbf{Statistical Code Completion} ~ 
Code completion is a hot research topic in the field of software engineering. Early work in code completion mainly bases on heuristic rules and static type information to make suggestions \cite{hou2010towards}. Since \citet{hindle2012naturalness} found that source code contained predictable statistical properties, statistical language models began to be used for modeling source code \cite{nguyen2013statistical,hellendoorn2017deep,Li2018Code,wei2019code,liu2020modeling}, where N-gram is the most widely used model. \cite{tu2014localness} observed that source code has a unique property of localness, which could not be captured by the traditional N-gram model. They improved N-gram by adding a cache mechanism to exploit localness and achieved better performance than other N-gram based models. \citet{hellendoorn2017deep} introduced an improved N-gram model that considered the unlimited vocabulary, nested scope, locality, and dynamism in source code. 

In recent years, deep recurrent neural network-based language models have been applied to learning source code and have made great progress \cite{liu2016neural,bhoopchand2016learning,Li2018Code,liu2019self,chen2020deep}. \citet{liu2016neural} proposed a code completion model based on a vanilla LSTM network. \citet{Li2018Code} proposed a pointer mixture network to address the OoV issue. \citet{liu2019self} propose a multi-task learning and transformer based language model for AST-level code completion. They built model to predict the AST node's type and value jointly and also utilized the hierarchical structural information in the program's representation, which achieves state-of-the-art results on AST-level code completion. \citet{kim2020code} presented a transformer model for code prediction and incorporated syntactic structure into the transformer to further improve the model's performance. \citet{svyatkovskiy2019pythia} proposed a code completion system based on LSTM for recommending Python method calls. Their system is deployed as part of the Intellicode extension in Visual Studio Code IDE. \citet{karampatsis2020big} proposed a large-scale open-vocabulary neural language model for source code, which leverages the BPE algorithm, beam search algorithm, and cache mechanism to both keep vocabulary size low and successfully predict OoV tokens. The experimental results demonstrate that their open vocabulary model outperforms both N-gram models and closed vocabulary neural language models, and achieve state-of-the-art performance on token-level code completion. Most recently, \citet{svyatkovskoy2020fast} implemented and evaluated a number of neural code completion models, which offer varying trade-offs in terms of memory, speed and accuracy. They provided a well-engineered approach to deep-learning based code completion, which is important to the software engineering community.

\noindent\textbf{Pre-trained Language Models}
Language model pre-training has shown to be effective for NLP, and has achieved the state-of-the-art results across many NLP tasks \cite{dai2015semi,Devlin2019Bert,Peters2018Deep,radford2018improving,Howard2018Universal}. Pre-trained language models can learn token contextualized representations by predicting tokens based on their context by training on large amounts of data, and then can be adapted to downstream tasks. Bidirectional Encoder Representations from Transformers (BERT) \cite{Devlin2019Bert} is the widely used approach in NLP, which learns to predict the masked words of a randomly masked word sequence given surrounding contexts. BERT has significantly improved the performance of a wide range of natural language understanding tasks. \citet{kanade2019pre} extended this idea to programming language understanding tasks. They derived contextual embedding of source code by training a BERT model on source code. They evaluate their model on a benchmark of five classification tasks in programs. Results show that their model outperforms the baseline LSTM models supported by Word2Vec embeddings, and Transformers trained from scratch. The bidirectionality nature of BERT makes it difficult to be applied to natural language generation tasks. To overcome this limitation, \citet{dong2019unified} proposed a unified pre-trained language model (UNILM) that can be applied to both natural language understanding and natural language generation tasks. UNILM can be configured using different self-attention masks to aggregate context for different types of language models, and thus can be used for both language understanding and generation tasks.

In the above work, the models are learned from the input of a single modal, for example, only from natural languages or source code. In recent years, multi-modal pre-trained models that can learn implicit alignment between inputs of different modalities are proposed. These models are learned from bi-modal data, such as pairs of language-image \cite{lu2019vilbert}, language-video \cite{sun2019videobert}, or language-code \cite{feng2020codebert}. \citet{feng2020codebert} proposed CodeBERT, a bimodal pre-trained model for natural language and programming language, aiming at capturing the semantic connection between natural language (NL) and programming language (PL). They trained CodeBERT with masked language modeling task and replaced token detection task, and evaluated it on two downstream NL-PL tasks, including natural language code search and code documentation generation.

Inspired by the above models, we propose a code understanding and generation pre-trained language model with a transformer-based architecture and tailored it for code completion, which is the first attempt at pre-training a language model for code completion.

\section{Conclusions and Future Work}\label{conclusion}
In this paper, we propose a multi-task learning based code understanding and generation pre-trained language model for source code modeling with a Transformer-based neural architecture. We pre-train our model on two massive datasets and with three objective functions and then fine-tune it on code completion. Experimental results demonstrate that the proposed model achieves better results than previous state-of-the-art models on completing tokens, especially on completing identifiers. To the best of our knowledge, we are the first to apply the pre-trained language model to code completion. We believe this work represents a significant advance in source code modeling, which will be beneficial as a building block for many other applications in this area. 

In the future, we plan to apply our model to other programming languages and fine-tune our model to adapt to other tasks.

\begin{acks}
This research is supported by the National Key R\&D Program under Grant No. 2018YFB1003904, and the National Natural Science Foundation of China under Grant Nos. 61832009, 61620106007, and 61751210.
\end{acks}

\bibliographystyle{ACM-Reference-Format}
\bibliography{ref}

\end{document}